\documentstyle[12pt]{article} 
\advance \textheight by \topskip 
\makeatletter 
\def\eqnarray{\stepcounter{equation}\let\@currentlabel=\theequation 
\global\@eqnswtrue 
\global\@eqcnt\z@\tabskip\@centering\let\\=\@eqncr 
$$\halign to \displaywidth\bgroup\@eqnsel\hskip\@centering 
  $\displaystyle\tabskip\z@{##}$&\global\@eqcnt\@ne  
  \hfil$\displaystyle{\hbox{}##\hbox{}}$\hfil 
  &\global\@eqcnt\tw@ $\displaystyle\tabskip\z@ 
  {##}$\hfil\tabskip\@centering&\llap{##}\tabskip\z@\cr} 
\@addtoreset{equation}{section} 
  \def\theequation{\thesection.\arabic{equation}} 
\makeatother 
\mathchardef\by="0202 
 
\begin{document} 

\thispagestyle{empty} 
\begin{flushright} 
{\bf NF/DF-05/96}\\ 
{\bf hep-th/9701091} 
\end{flushright} 
$\ $ 
\vskip 2truecm 
\begin{center} 
{ \Large \bf Deformed Heisenberg algebra with reflection}\\
\vskip1.0cm 
{ \bf Mikhail S. Plyushchay{}\footnote{ 
E-mail: plushchay@mx.ihep.su}}\\[0.3cm] 
{\it Departamento de F\' {i}sica -- ICE}\\ 
{\it Universidade Federal de Juiz de Fora}\\ 
{\it 36036-330 Juiz de Fora, MG Brazil}\\ 
{\it and}\\ 
{\it Institute for High Energy Physics}\\ 
{\it Protvino, Moscow Region, 142284 Russia}  
\end{center} 
\vskip2.0cm 
\begin{center}                             
{\bf Abstract} 
\end{center} 
A universality of deformed Heisenberg algebra involving the
reflection operator is revealed.  It is shown that in addition
to the well-known infinite-dimensional representations related
to parabosons, the algebra has also finite-dimensional
representations of the parafermionic nature.  We demonstrate
that finite-dimensional representations are representations of
deformed parafermionic algebra with internal $Z_2$-grading
structure.  On the other hand, any finite- or
infinite-dimensional representation of the algebra supply us
with irreducible representation of $osp(1|2)$ superalgebra.  We
show that the normalized form of deformed Heisenberg algebra
with reflection has the structure of guon algebra related to the
generalized statistics. 
\vskip0.3cm 
{\bf Key words:} deformed Heisenberg algebra, parabosons, 
deformed parafermionic algebra, 
$osp(1|2)$ superalgebra, generalized statistics 
\vskip1.0cm 
\begin{center} 
To appear in {\it Nucl. Phys.} {\bf B}  
\end{center} 
\newpage

\section{Introduction}
The deformed Heisenberg algebra involving the reflection
operator $R$ appeared in the context of quantization schemes
generalizing bosonic commutation relations.  Such generalized
schemes lead naturally to the concept of parafields and
parastatistics \cite{para,ok,mac1}.  The same 
$R$-deformed Heisenberg algebra (RDHA)
was also used for solving quantum mechanical Calogero model
\cite{yang,calog,bem}. Recently this algebra has been employed 
for bosonization of supersymmetric quantum mechanics 
\cite{bem,bosany,bos} and for describing anyons 
in (2+1) \cite{bosany,any} and (1+1) dimensions \cite{any1}.  
All the applications as well as the parabosonic constructions
\cite{ok,mac1} use infinite-dimensional unitary representations of 
RDHA.

In this paper we shall reveal a universality of the $R$-deformed
Heisenberg algebra.

We shall show that in addition to the infinite-dimensional
representations related to parabosons, the algebra has also
finite-dimensional representations of the parafermionic nature.
Finite-dimensional representations will be investigated in
different aspects.  This includes their interpretation as
representations of some non-degenerate paragrassmann algebra
\cite{fik1,fik2} as well as treating them as representations of
generalized deformed parafermionic \cite{que} and 
deformed $su(2)$ algebras \cite{polroc}.  
We shall find that any irreducible
representation of RDHA supplies us with corresponding finite- or
infinite-dimensional irreducible representation of $osp(1|2)$
superalgebra.  The $osp(1|2)$ generators turn out to be
realizable in a universal form in terms of generators of the
underlying deformed Heisenberg algebra. Besides, we shall show
that the normalized form of RDHA has the structure of guon
algebra \cite{guon} related to generalized statistics
\cite{stat,genon}.

The paper is organized as follows.  In Section 2 we consider
Fock space representations of the $R$-deformed Heisenberg
algebra.  Here the relationship of the algebra
to parabosonic trilinear (anti)commutation relations is
discussed and finite-dimensional representations of
parafermionic type are found.  Finite-dimensional
representations are analyzed in detail in Section 3, where we
discuss also a universal $osp(1|2)$ superalgebraic aspect of
RDHA.  Section 4 is devoted to the construction of the
normalized form of the algebra.  
We get the standard fermionic algebra, presented in the
form of $R$-algebra, as a limit case of the normalized RDHA and
generalize it into the algebra with phase operator.  We show that
such a generalization is  related to the $q$-deformed
Heisenberg algebra \cite{qdef,macbi} with $q$ being a primitive
root of unity.  In Section 5 we discuss possible applications of
the results.

Appendix A concerns the realization of finite-dimensional
representations on the $q$-paragrassmann algebra
\cite{fik1,fik2}.

Appendix B illustrates a relationship of two lower-dimensional
representations of RDHA to the spinor and vector representations
of (2+1)-dimensional Lorentz group.

\vskip0.5cm

The paper is devoted to the memory of Dmitrij Vasilievich Volkov.

\section{Representaions of $R$-deformed Heisenberg algebra}

The $R$-deformed Heisenberg algebra is given by the generators 
$1$, $a^-$, $a^+$, $R$, which 
satisfy the (anti)commutation relations
\begin{equation}
[a^-,a^+]=1+\nu R,\quad 
\{a^\pm,R\}=0,\quad R^2=1,
\label{rdef}
\end{equation}
and $[a^\pm,1]=[R,1]=0$. Here $\nu\in {\bf R}$ is a deformation
parameter and hermitian operator
$R$ is the reflection operator \cite{ok}.
The relationship of $a^+$ and $a^-$ under hermitian
conjugation will be specified lately.

As a consequence of eq. (\ref{rdef}),
operators $a^\pm$ obey trilinear
(anti)commutation relations not containing either operator $R$
or deformation parameter:
\begin{equation}
[\{a^-,a^+\},a^-]=-2a^-,\quad
[\{a^-,a^+\},a^+]=2a^+.
\label{parabo}
\end{equation}
Trilinear relations (\ref{parabo})
characterize parabosons \cite{para,ok}.
They can be represented equivalently as 
$[a^+,a^{-2}]=-2a^-$ and $[a^-,a^{+2}]=2a^+$, or as
\begin{equation}
\{a^-,[a^-,a^+]\}=2a^-,\quad
\{a^+,[a^-,a^+]\}=2a^+.
\label{parabo*}
\end{equation} 
These trilinear relations  themselves
lead to $R$-deformed Heisenberg algebra (\ref{rdef}).
To show this, let us define operator
$G=[a^-,a^+]-1$.
Relations (\ref{parabo*}) mean
that $\{G,a^\pm\}=0$ and as a consequence, $[G^2,a^\pm]=0$.
Hence, for irreducible representation 
we have $G^2=const$. If in such representation 
$G$ is hermitian operator,
then $G=\nu R$ with  $\{R,a^\pm\}=0$, $R^2=1$ and
$\nu\in{\bf R}$. 

Now we proceed to consideration of irreducible representations
of algebra (\ref{rdef}).
One introduces the vacuum state $|0\rangle$,
$a^-|0\rangle=0$, $\langle0|0\rangle=1$, $R|0\rangle=|0\rangle$,
and defines the states $|n\rangle=C_n (a^+)^n|0\rangle$, $n=0,1,\ldots,$
with $C_n$ being some normalization constants.
{}From the relation
\begin{equation}
[a^-,(a^+)^n]=
\left(n+\frac{1}{2}(1-(-1)^n)\nu R\right)(a^+)^{n-1}
\label{rn}
\end{equation}
we conclude that algebra (\ref{rdef})
has infinite-dimensional unitary representations when 
$
\nu>-1. 
$
Only in this case the states 
$|n\rangle$ with $C_n=([n]_\nu !)^{-1/2}$, 
$[n]_\nu !=\prod_{l=1}^n [l]_\nu$, 
$[l]_\nu=l+\frac{1}{2}(1-(-1)^l)\nu$,
form the complete orthonormal basis of Fock representation
and operators $a^+$ and $a^-$ are hermitian conjugate.
The reflection operator can be realized in terms of 
creation and annihilation operators via the number 
operator $N$,
\begin{equation}
R=(-1)^N=\cos\pi N,
\label{RN}
\end{equation}
\begin{equation} 
N=\frac{1}{2}\{a^+,a^-\}-\frac{1}{2}(\nu+1),\quad
N|n\rangle=n|n\rangle.
\label{NN}
\end{equation}

Trilinear commutation relations (\ref{parabo}) characterize the
{\it parabosonic} system of order $p=1,2,\ldots$ in the case
when $\nu=p-1=0,1,\ldots$ \cite{para,ok}.  
The existence of infinite-dimensional unitary representations
of algebra (\ref{rdef}) on the half-line
$\nu>-1$ and the relationship between trilinear commutation relations and
$R$-deformed Heisenberg algebra (\ref{rdef}) mean that the
latter can be considered as the algebra supplying us with some
generalization of parabosons for the case of non-integer
statistical parameter $p=\nu+1>0$.  
Such generalized parabosonic aspect 
of algebra (\ref{rdef}) was discussed recently by
Macfarlane \cite{mac1}.

Now we shall show that algebra (\ref{rdef}) 
has finite-dimensional representations,
and, as a consequence, it is also related to 
{\it parafermionic} systems. To this end we note  
that eq. (\ref{rn}) gives 
some special values of the deformation parameter, 
\[
\nu=-(2p+1),\quad
p=1,2,\ldots,
\]
for which the relation
$
\langle\langle m|n\rangle\rangle=0,
$
$|n\rangle\rangle\equiv(a^+)^n|0\rangle$,
takes place for $n\geq 2p+1$ and arbitrary $m$. 
This means that there are $(2p+1)$-dimensional irreducible
representations of algebra (\ref{rdef}) with
$\nu=-(2p+1)$, in which the relations $(a^+)^{2p+1}=(a^-)^{2p+1}=0$
are valid. The latter relations
are a characteristic property of parafermions of order $2p$
\cite{ok}.
The relationship of these $(2p+1)$-dimensional representations
of the $R$-deformed Heisenberg algebra 
to parafermions will be discussed in detail in 
the next section.

One  concludes that the $R$-deformed Heisenberg algebra 
turns out to be related 
to parabosons and to parafermions.
In this sense it has some properties of universality.

\section{Finite-dimensional representations} 
We shall investigate finite-dimensional
representations in different aspects.  First we shall discuss
them as representations of some non-degenerate paragrassmann
algebra.  Then we shall obtain the corresponding matrix
representation.  In particular, we shall show that operators
$a^+$ and $a^-$ are mutually conjugate with respect to
indefinite scalar product. We shall find another set of
operators, $f^+$ and $f^-$, which are related
to $a^+$ and $a^-$ in a simple way.
These new operators are hermitian and give a possibility to
work with a Hilbert space of states.  Operators $f^\pm$,  unlike
$a^\pm$, will satisfy {\it anticommutation} relations involving
the reflection operator.  We shall show that 
$f^\pm$ are the creation-annihilation operators of
some deformed parafermionic algebra, which at $p=1$ is reduced
to the standard parafermionic algebra of order 2.  
For $p>1$, operators
$f^\pm$ generate the related special deformation of
$su(2)$ algebra. 
The realization of deformed $su(2)$ generators in terms 
of the standard $su(2)$ generators
reflects a nontrivial relationship of our
deformed parafermionic algebra to the standard parafermionic
algebra of the same order.
We shall show that there is also a simple 
realization of the standard $su(2)$ generators in terms of
operators $f^\pm$.  However, the price necessary to pay 
will be reducible action of these $su(2)$ generators.  
The related reducible 
quadratic realization of $so(2,1)$
will be obtained in terms of initial operators $a^\pm$.
It will give the standard (non-unitary) finite-dimensional
representations of $(2+1)$-dimensional 
Lorentz group.  The above-mentioned indefinite scalar product
turns out to be the standard indefinite scalar product which is
necessary to have Lorentz generators as self-conjugate
operators.
We shall show that operators $a^+$ and $a^-$ together with 
quadratic Lorentz generators 
form the set of generators of $osp(1|2)$ superalgebra
and give us its $(2p+1)$-dimensional irreducible representations.

\subsection{R-paragrassmann algebra}
Let us consider finite-dimensional representations in more detail.
We have arrived at the nilpotent algebra
\begin{eqnarray}
&[a^-,a^+]=1-(2p+1)R,\quad
\{a^\pm,R\}=0,\quad R^2=1,&
\label{aaa}\\
&(a^\pm)^{2p+1}=0,\quad
p=1,2,\ldots.&
\label{parag}
\end{eqnarray}
As a consequence of relations (\ref{aaa}), (\ref{parag}),
we have the relations
\begin{equation}
(1-R)a^{+2p}=(1-R)a^{-2p}=0.
\label{pauli}
\end{equation}
These relations and eqs. (\ref{aaa}), in turn, lead to conditions 
(\ref{parag}). This means that here
the nilpotency conditions (\ref{parag}) are
equivalent to relations (\ref{pauli}).

One can interpret $a^+$ as a
paragrassmann variable $\theta$, $\theta^{2p+1}=0$.
Then operator
$a^-$ can be considered as corresponding differentiation operator 
$\partial$
defined by relation (\ref{rn}) (see refs. \cite{fik1,fik2}). 
Therefore, algebra (\ref{aaa}), (\ref{parag}) is a paragrassmann
algebra of order $2p$ with special differentiation operator.
One can call it the R-paragrassmann algebra.

In addition to the universal realization (\ref{RN}), (\ref{NN}),
for the $R$-paragrassmann algebra
we have the normal ordered representation 
\begin{equation}
R=\sum_{n=0}^{2p}f_{n}a^{+n}a^{-n}.
\label{rnor}
\end{equation}
The $c$-number factors $f_n$ are given 
by finite recursive relations 
\[
2f_{n-1}+[n]_\nu f_n-(2p+1)\sum_{i=0}^{[n/2]-1}f_{2i+1}
f_{n-(2i+1)}=0,\quad
n=1,\ldots, 2p,
\]
where $[n/2]$ is an integer part
of $n/2$, and as it follows just from eq. (\ref{pauli}), $f_0=1$.
In simplest cases this gives
\[
R=1+a^+a^-+\frac{1}{2}a^{+2}a^{-2},\quad
p=1,
\]
\[
R=1+\frac{1}{2}a^+a^-+\frac{1}{8}a^{+2}a^{-2}
-\frac{1}{32}a^{+3}a^{-3}-\frac{3}{128}a^{+4}a^{-4},\quad
p=2.
\]

The explicit form of normal ordered representation (\ref{rnor})
and the form of commutation relations (\ref{rn}) 
say that the $R$-paragrassmann algebra is non-degenerate 
paragrassmann algebra 
characterized by the properties 
$\partial\theta^n=\alpha_n\theta^{n-1}+(...)\partial$,
$\alpha_n\neq 0,$ $n=1,2,\ldots,p'$,
$\theta^{p'+1}=0$ \cite{fik2}. Due to isomorphism between 
non-degenerate paragrassmann algebras of the same order \cite{fik2},
it is possible to realize algebra (\ref{aaa}), 
(\ref{parag}) on the $q$-paragrassmann algebra with $q$ being 
a corresponding primitive root of unity. 
Such a realization is given in Appendix A.

\subsection{Matrix realization}
Finite-dimensional Fock space
representations of $R$-deformed Heisenberg algebra (\ref{rdef})
contain states with negative norm. 
One can introduce the normalized states as
$|n\rangle=|\langle\langle
n|n\rangle\rangle|^{-1/2}\cdot|n\rangle\rangle$.
They give the indefinite metric operator $\eta=\eta^\dagger$,
$\eta^2=1$, with matrix
elements 
\[
||\eta||_{mn}=\langle
m|n\rangle=diag(1,-1,-1,+1,+1,-1,-1,\ldots,
(-1)^{p-1},(-1)^{p-1},(-1)^p,(-1)^p).
\] 
Then the indefinite scalar product can be defined as 
\begin{equation}
(\Psi_1,\Psi_2)=\langle\Psi_1|\eta\Psi_2\rangle=
\Psi^*_{1n}\eta_{nm}\Psi_{2m}, 
\label{scapro}
\end{equation}
where $\Psi_n=\langle n|\Psi\rangle$.
Operators $a^+$ and $a^-$ are given by 
the matrices
\begin{equation}
(a^+)_{mn}=A_n\delta_{m-1,n},\quad
(a^-)_{mn}=B_m\delta_{m+1,n},
\label{mat1}
\end{equation}
with
\begin{equation}
A_{2k+1}=-B_{2k+1}=\sqrt{2(p-k)},\quad
k=0,1,\ldots,p-1,\quad
A_{2k}=B_{2k}=\sqrt{2k},\quad
k=1,\ldots,p.
\label{mat2}
\end{equation}
They satisfy relations (\ref{rdef}) with diagonal operator  
$R=diag(+1,-1,+1,\ldots,-1,+1)$.
Due to the relation 
$(a^-)^\dagger=Ra^+=\eta a^+\eta$,
operators $a^+$ and $a^-$ are mutually conjugate with respect
to indefinite scalar product
(\ref{scapro}), $(\Psi_1,a^-\Psi_2)^*=
(\Psi_2,a^+\Psi_1)$. 

One can introduce hermitian conjugate operators 
\begin{equation}
f^+=a^+,\quad f^-=a^-R
\label{ff}
\end{equation}
instead of operators $a^\pm$.
Then finite-dimensional representations of the $R$-deformed
Heisenberg algebra can be specified by the relations
\begin{eqnarray}
&&\{f^+,f^-\}=(2p+1)-R,
\label{ff*}\\
&&\{R,f^\pm\}=0,\quad
R^2=1,\quad
(f^\pm)^{2p+1}=0.
\label{fff}
\end{eqnarray}
{}From now on one could work 
in terms of hermitian conjugate operators (\ref{ff})
using ordinary positive-definite scalar product 
$(\Psi_1,\Psi_2)=\Psi^*_{1n}\Psi_{2n}$.
Operators $a^+=f^+$, $a^-=f^-R$ can be considered 
in this case as not basic operators. 
However, we shall see further that the usage of operators 
$a^+$, $a^-$ as basic conjugate operators 
together with corresponding indefinite scalar product turns out
also to be useful for some physical applications.

The present situation with two sets of operators, $a^+$, $a^-$
and $f^+$, $f^-$, is similar to that
taking place for finite-dimensional representations
of the $q$-deformed Heisenberg algebra \cite{qdef,macbi}.
That algebra can be given by commutation 
relations in the Biedenharn-Macfarlane form
$b^-b^+-qb^+b^-=q^{-N}$ with $N$ being a number operator. 
In the case $q^{k+1}=-1$, $k=1,2,\ldots$,
the algebra has $(k+1)$-dimensional representations 
with $(b^\pm)^{k+1}=0$,
in which operators $b^+$ and $b^-$ are hermitian conjugate.
In terms of operators $c^+=b^+ q^{N/2}$ and $c^-=q^{N/2}b^-$,
the algebra is given in more simple Arik-Coon-Kuryshkin form,
$c^-c^+-q^2 c^+c^-=1$, but here operators 
$c^\pm$ are not hermitian conjugate, $(c^+)^\dagger=q^{-N}c^-$.

\subsection{Parafermions and deformed $su(2)$}
Here we shall show that finite-dimensional representations of
the $R$-deformed Heisenberg algebra are representations of
deformed parafermionic algebra which is related to some
nonlinear deformation of $su(2)$.  In the case of lowest
representation ($p=1$), the deformed parafermionic algebra is
reduced to the standard parafermionic algebra of order $2$
giving us a vector representation of the standard $su(2)$.

{}First one notes that the operators $a^\pm$ satisfying
eqs. (\ref{aaa}), (\ref{parag}) obey the relations
\begin{eqnarray}
&a^+a^{-2p}+a^-a^+a^{-(2p-1)}+\ldots 
+a^{-(2p-1)}a^+a^-+a^{-2p}a^+=0,&\nonumber\\
&a^-a^{+2p}+a^+a^-a^{+(2p-1)}+\ldots 
+a^{+(2p-1)}a^-a^++a^{+2p}a^-=0.&
\label{parasusy}
\end{eqnarray}
Such relations together with relations
$a^{\pm(2p+1)}=0$ take place for 
$N=2$ parasupercharge operators 
in the case of trivial parasupersymmetry 
characterized by zero parasuperhamiltonian \cite{rubspi}.
The alternating sum 
\begin{equation}
f^+f^{+2p}-f^-f^+f^{-(2p-1)}+\ldots
-f^{-(2p-1)}f^+f^-+f^{-2p}f^+=0
\label{ffp}
\end{equation}
and its hermitian conjugate analog are
equivalent to relations (\ref{parasusy}).
The following relations are also valid:
\[
f^{+p}f^-f^{+p}=f^{+(2p-1)}C_p,\quad
f^{-p}f^+f^{-p}=f^{-(2p-1)}C_p,
\]
where $C_p=-p$ for even $p$ and $C_p=p+1$ for odd $p$.
At $p=1$ they are reduced to
\begin{equation}
f^-f^+f^-=2f^-,\quad f^+f^-f^+=2f^+.
\label{para1}
\end{equation}
Taking into account eq. (\ref{para1}),
we reduce equation (\ref{ffp}) and its conjugate 
to the relations
\begin{equation}
f^{-2} f^++f^+f^{-2}=2f^-,\quad
f^-f^{+2}+f^{+2}f^-=2f^+.
\label{para2}
\end{equation}
As a consequence of eqs. (\ref{para1}), (\ref{para2}),
at $p=1$ one has
\begin{equation}
[[f^+,f^-],f^+]=2f^+,\quad
[[f^+,f^-],f^-]=-2f^-.
\label{trif}
\end{equation}
Relations (\ref{para1}), (\ref{para2}) and $f^{\pm3}=0$
are the relations characterizing 
the parafermions of order 2.
Spin-1 $su(2)$ generators 
are realized in terms of these parafermionic operators as 
$I_+=f^+,$
$I_-=f^-,$
$I_3=\frac{1}{2}[f^+,f^-].$
For $p>1$, operators $I_+=f^+,$
$I_-=f^-$
generate a nonlinear deformation of $su(2)$ algebra of the form
\begin{eqnarray}
&&[I_+,I_-]=2I_3(-1)^{I_3+p},
\label{ffi3}\\
&&[I_3,I_\pm]=\pm I_\pm,
\label{su2d}
\end{eqnarray}
involving the reflection operator
\begin{equation}
R=(-1)^{I_3+p}.
\label{ri}
\end{equation}
At $p=1$, one has the relation $I_3(-1)^{I_3+1}=I_3$, and deformed
$su(2)$ algebra (\ref{ffi3}),
(\ref{su2d}) turns into the standard $su(2)$.
Relations (\ref{ffi3}),
(\ref{su2d}) can be represented in the form 
\begin{equation}
[[f^+,f^-],f^\pm]=2(2I_3\mp 1)(-1)^{I_3+p}f^\pm
\label{dfp}
\end{equation}
with operator $I_3$ given by 
\begin{equation}
I_3=\frac{1}{2}[f^+,f^-]\cdot(-1)^{\frac{1}{2}[f^+,f^-]+p}.
\label{i3}
\end{equation}
Relations (\ref{dfp}) generalize trilinear parafermionic relations
(\ref{trif}) for the case $p>1$ and mean that we have deformed
parafermionic algebra.

The (anti)commutation relations (\ref{ff*}) 
and (\ref{ffi3}) can be represented as
\begin{equation}
\{f^+,f^-\}=F(N+1)+F(N),\quad
[f^-,f^+]=F(N+1)-F(N),
\label{gd1}
\end{equation}
with function $F(N)=N(-1)^N+(p+\frac{1}{2})(1-(-1)^N)$
being a function of the number operator $N=I_3+p$.  
This function is characterized by the properties
\begin{equation}
F(n)>0,\quad n=1,2,\ldots,p',
\quad
F(p'+1)=0,
\label{gd2}
\end{equation}
and here $p'=2p$.
The algebra with defining relations  (\ref{gd1}), (\ref{gd2}) is
the generalized deformed parafermionic algebra introduced by
Quesne \cite{que}. 
For $F(x)=x(p'+1-x)$ it is reduced to the
standard parafermionic algebra of order $p'$.  
It is interesting to note that
algebra (\ref{gd1}), (\ref{gd2}) contains also $q$-deformed
parafermionic algebra \cite{deffq}, which for $p'=2$
is reduced to the standard parafermionic algebra
as it happens in our case.

Generators of deformed $su(2)$ algebra (\ref{ffi3}), (\ref{su2d})
can be related to the generators 
of the standard $su(2)$ algebra,
\begin{equation}
[J_+,J_-]=2J_3,\quad
[J_3,J_\pm]=\pm J_\pm,
\label{stsu2}
\end{equation}
as
\begin{equation}
I_3=J_3,\quad I_-=J_-\Phi(J_3),\quad
I_+=\Phi(J_3)J_+.
\label{ijd}
\end{equation}
Here 
function $\Phi(J_3)$ is given by 
\begin{equation}
\Phi(J_3)=\sqrt{\frac{2p+1+R(2J_3-1)}{2(p+J_3)(p-J_3+1)}},\quad
R=(-1)^{J_3+p}.
\label{defphi}
\end{equation}
We assume that generators $J_\pm$, $J_3$
realize 
$(2p+1)$-dimensional representation of $su(2)$ 
which are characterized by
the corresponding value of the
Casimir operator, $J_3^2+\frac{1}{2}\{J_+,J_-\}=p(p+1)$.
Definition (\ref{defphi})
gives indefinite relation of the form $0/0$ at $J_3=-p$, where
it should be supplemented by any finite nonzero value. 
This supplementing a definition
is formally necessary to have a well defined action 
of operator $I_-$ ($I_+$) on the state 
$|-p\rangle$ ($\langle -p|$) being the 
eigenstate of $J_3$, $J_3|-p\rangle=-p|-p\rangle$.
Then, for $p>1$,
the transformation inverse to (\ref{ijd}) and eq. (\ref{i3})
give essentially nonlinear realization of the standard
$su(2)$ in terms of operators $f^\pm$.
Further we shall show that the
special nature of our deformed 
parafermionic algebra admits, nevertheless, a simple 
(but reducible) realization
of the standard $su(2)$.
Note here that 
the deformations of $su(2)$  of the form (\ref{ijd})
were considered by Polychronakos and Ro\v{c}ek \cite{polroc}.

\subsection{Finite-dimensional representations of $osp(1|2)$}.
Our deformed parafermionic algebra
admits also a simple realization of the standard $su(2)$. 
This turns out to be possible due to a special $Z_2$-grading 
structure of deformed $su(2)$ algebra (\ref{ffi3}), (\ref{su2d}),
which consists in the presence of reflection operator (\ref{ri}) in
deformed commutation relation (\ref{ffi3}).
The related construction 
in terms of initial operators $a^\pm$
will supply us with quadratic realization of generators
of (2+1)-dimensional Lorentz group
and will give a self-conjugate set of $osp(1|2)$
generators. As we shall see, the indefinite scalar product
introduced in section 3.2 will be recognized 
as the standard scalar product necessary 
to have spin-$j$ Lorentz 
generators as self-conjugate operators.

One introduces hermitian conjugate quadratic operators 
\begin{equation}
J_+=\frac{1}{2}f^{+2},\quad
J_-=\frac{1}{2}f^{-2}.
\label{j+j-}
\end{equation}
Direct calculation gives 
$
[J_+,J_-]=2J_3,
$
with
\begin{equation}
J_3=\frac{1}{2}I_3,
\label{jtri}
\end{equation}
where $I_3$ is the diagonal operator introduced before,
$I_3=N-p$. Eq. (\ref{su2d}) leads to the relations
$
[J_3,J_\pm]=\pm J_\pm.
$
Therefore, operators (\ref{j+j-}) and (\ref{jtri})
form, unlike the  
generators $I_\pm$, $I_3$, the standard $su(2)$ algebra.
As a consequence, operators (\ref{j+j-})
satisfy standard trilinear parafermionic
commutation relations of the form (\ref{trif}).
But since $(f^{\pm2})^{p+1}=0$, our $(2p+1)$-dimensional representation
is reducible with respect to the action of 
operators (\ref{j+j-}) as parafermionic generators. 
It is a direct sum of $(p+1)$-dimensional even 
and of $p$-dimensional odd  subspaces
spanned by  even, $|+\rangle$, and odd, $|-\rangle$, 
eigenstates of reflection operator, 
$R|\pm\rangle=\pm|\pm\rangle$. These subspaces carry 
unitary spin-$j_+$ 
and spin-$j_-$  $su(2)$-representations: 
$C|+\rangle=j_+(j_++1)$,
$C|-\rangle=j_-(j_-+1)$, where $C$ is the $su(2)$ quadratic 
Casimir operator,
$C=J_3^2+\frac{1}{2}\{J_+,J_-\}$, and
$j_+=p/2$, $j_-=(p-1)/2$. 

Let us consider operators ${\cal J}_+=J_+$, 
${\cal J}_-=-J_-$, ${\cal J}_0=J_3=\frac{1}{4}[f^+,f^-]R$.
In terms of operators $a^\pm$,
they are represented as quadratic operators:
\begin{equation}
{\cal J}_0=\frac{1}{4}\{a^+,a^-\},\quad
{\cal J}_+=\frac{1}{2}a^{+2},\quad
{\cal J}_-=\frac{1}{2}a^{-2}.
\label{osp12}
\end{equation}
These generators form $so(2,1)$ algebra,
\begin{equation}
[{\cal J}_0,{\cal J}_\pm]=\pm{\cal J}_\pm,
\quad
[{\cal J}_-,{\cal J}_+]=2{\cal J}_0.
\label{so21}
\end{equation}
Such realization gives reducible representation
of this algebra being the direct sum of spin-$j_+$
and spin-$j_-$ representations:
${\cal C}|\pm\rangle=-j_\pm(j_\pm+1)|\pm\rangle$,
where 
${\cal C}=-{\cal J}_0^2+\frac{1}{2}\{{\cal J}_+,{\cal J}_-\}$
is $so(2,1)$ Casimir operator, whereas 
subspaces $|\pm\rangle$ and numbers $j_\pm$
have been specified above.
Operators $a^+$ and $a^-$, being `square root' operators 
of $so(2,1)$ generators, have with them the following 
nontrivial commutation relations:
\begin{equation}
[{\cal J}_+,a^-]=-a^+,\quad
[{\cal J}_-,a^+]=a^-,\quad
[{\cal J}_0,a^\pm]=\pm\frac{1}{2}a^\pm.
\label{jaa}
\end{equation}
These relations mean that operators $a^+$ and $a^-$ are components 
of $(2+1)$-dimensional spinor. 
Reading relations (\ref{osp12}) in inverse order,
from the right to the left hand side,
and taking into account eqs. (\ref{so21}), (\ref{jaa}),
one concludes that operators $a^+$, $a^-$, ${\cal J}_0$, 
${\cal J}_+$ and ${\cal J}_-$ form $osp(1|2)$ superalgebra
\cite{susy}
with $a^\pm$ being odd generators and 
${\cal J}_0$, ${\cal J}_\pm$ being even generators of the
superalgebra. The corresponding Casimir operator
of the superalgebra is $C_{osp}=-{\cal J}_0^2+\frac{1}{2}\{{\cal J}_+,
{\cal J}_-\}-\frac{1}{8}[a^-,a^+]$, and it takes here
the value $C_{osp}=-\frac{1}{4}p(p+1)$.
Therefore, every $(2p+1)$-dimensional representation of the $R$-deformed
Heisenberg algebra supplies us with irreducible representation
of the standard (non-deformed) $osp(1|2)$ superalgebra.
To have operators $a^+$, $a^-$ and ${\cal J}_+$, ${\cal J}_-$
as mutually conjugate operators, it is necessary to use indefinite 
scalar product (\ref{scapro}).
There is nothing surprising in this fact since it is well known
that finite-dimensional representations of $so(2,1)$ 
are non-unitary (see, e.g., ref. \cite{corply}). 
Dirac conjugate field $\bar{\psi}$
containing $\gamma^0$-factor in the case of spinor representation
and indefinite metric operator 
of the form $\eta_{\mu\nu}=diag(-1,+1,+1)$ 
in the case of vector representation 
reflect this non-unitarity. 
The relationship of indefinite scalar product 
(\ref{scapro}) to the above mentioned standard simplest physical
manifestations of non-unitarity of finite-dimensional $so(2,1)$
representations is illustrated in Appendix B.

Therefore, we have shown that $(2p+1)$-dimensional
representations of the $R$-deformed Heisenberg algebra supply us
with irreducible (non-unitary) representations of $osp(1|2)$
superalgebra.  They are analogs of the well known
infinite-dimensional half-bounded unitary representations of
this superalgebra, which are realized on corresponding
infinite-dimensional representations of the $R$-deformed
Heisenberg algebra.  In the case of infinite-dimensional
representations the superalgebra generators are realized via
operators $a^\pm$ in the same form  as here (see refs.
\cite{bem,bosany,bos}).  Such universality of the construction
comprising finite-dimensional and infinite-dimensional
representations of $osp(1|2)$ can find interesting application
in (2+1)-dimensional physics.  This point will be discussed in
the last section.

\section{Normalized $R$-deformed Heisenberg algebra}

We have analyzed representations of the $R$-deformed Heisenberg
algebra proceeding from its standard non-normalized form (\ref{rdef}).
Here we shall find the normalized form of the algebra.
Using such a terminology we 
have in mind the normalization of the right hand
side of the corresponding commutation relations.
As we shall see, the normalized form represents 
by itself a guon-like algebra \cite{guon}
which is similar to a normalized form of $q$-deformed
Heisenberg algebra, $c^-c^+-qc^+c^-=1$ \cite{qdef},
but unlike the latter, it
contains some special $g$-operator factor instead of $c$-number
$q$-factor. The initial form of our algebra
(\ref{rdef}) corresponds to the non-normalized
variant of the $q$-deformed Heisenberg algebra, 
$b^-b^+-qb^+b^-={q}^{-N}$ \cite{macbi}.

\subsection{Guons} 
Let us suppose that $\nu\neq1$, and define
the operators 
\begin{equation}
c^-=a^-G_\nu^{-1/2}(R),\quad
c^+=G_\nu^{-1/2}(R) a^+,\quad 
G_\nu(R)=|1-\nu R|,
\label{car}
\end{equation}
where for the moment we suppose that $R=(-1)^N$
with $N$ given by eq. (\ref{NN}).
These operators anticommute with the
reflection operator, $\{R,c^\pm\}=0$, and satisfy the
commutation relation
$
c^-c^+-G_\nu(R) G_{\nu}^{-1}(-R)c^+ c^-=sign(1+\nu R),
$
where $sign\, x$ is $+1$ for $x>0$ and $-1$ for $x<0$.
The operator $G_\nu(R)$ in transformation (\ref{car}) 
is reduced to
$G_\nu(R)=1-\nu R$ for 
$-1<\nu<1$; for two other cases we have
$G_\nu(R)=\nu-R$,
$\nu>1$, and
$G_\nu(R)=R-\nu$, 
$\nu=-(2p+1)$.
As a result,
commutation relation is represented in first case in a normalized form
\begin{equation}
c^- c^+-g_{\nu} c^+c^-=1,\quad
g_\nu=(1-\nu)^R(1+\nu)^{-R},
\quad
-1<\nu<1,
\label{1nu1}
\end{equation}
whereas in two other cases it is reduced to
\begin{equation}
c^- c^+-g_{\nu}c^+c^-=R,
\label{rfer}
\end{equation}
where $g_\nu=(\nu-1)^R(1+\nu)^{-R}$ for $\nu>1$,
and $g_\nu=p^R(p+1)^{-R}$ for $\nu=-(2p+1)$.
In the case corresponding to finite-dimensional representations,
the final form (\ref{rfer}) has been obtained via additional
changing $R\rightarrow -R$.  In all three cases operator-valued
function $g_\nu$ satisfies the relation $g_\nu c^\pm=c^\pm
g_\nu^{-1}$.  The normalized form of algebra (\ref{1nu1}) is
similar to the Scipioni guon algebra \cite{guon} which was
introduced in the context of generalized statistics \cite{stat,genon}.
Algebra (\ref{rfer}) gives some modification of (\ref{1nu1}).

The corresponding number operator $N=N(c^+,c^-)$ is given by 
\[
N=-\frac{\alpha}{2}+\frac{1}{2}\sqrt{|1-\nu^2|
(2c^+c^--\beta)
(2c^-c^+-\beta)+1},
\]
where 
$\alpha=-\nu$, $\beta=1$ in the case $-1<\nu<1$,
and $\alpha=-\nu+\nu^2-1$, $\beta=|\nu|$
for two other cases. 
Implying in relations (\ref{1nu1}), (\ref{rfer}) 
that  $R=(-1)^{N(c^+,c^-)}$, one 
can represent them in a closed form containing only
creation-annihilation operators $c^\pm$.

\subsection{Fermions and $P$-algebra}
Let us consider a limit 
$|\nu|\rightarrow \infty$
proceeding from relation (\ref{rfer}).
Both cases, $\nu>-1$ and $\nu=-(2p+1)$, lead to the algebra
\begin{equation}
c^-c^+-c^+ c^-=R,\quad \{R,c^\pm\}=0,\quad R^2=1.
\label{antir}
\end{equation}
(Anti)commutation relations (\ref{antir}) 
have irreducible two-dimensional Fock space 
representation in which 
$c^-|0\rangle=0$, $R|0\rangle=|0\rangle$ and
$(c^\pm)^2=0$.
This corresponds to
fermionic representation.  Indeed, here
reflection operator is realized as $R=1-2c^+c^-$, and 
$R$-commutation relations (\ref{antir}) 
are reduced to the fermionic relations
$
c^+c^-+c^+c^-=1,
$
$
c^{+2}=c^{-2}=0.
$
Therefore, fermionic algebra
can be obtained from normalized $R$-deformed Heisenberg algebra
(\ref{rfer}) in the limit
$|\nu|\rightarrow+\infty$.

Algebra (\ref{antir}) has a natural
generalization related to the $q$-deformed Heisenberg algebra.
To show this, we note that $R$ is a phase operator, $R^2=1$. 
Then commutation relation (\ref{antir}) can be generalized to
\begin{equation}
[a,\bar{a}]=P,
\label{algp}
\end{equation}
where $P$ is a phase operator with the properties
\begin{equation}
P^{p'}=1,\quad 
Pa=q aP,\quad
P\bar{a}=q^{-1}\bar{a}P,\quad
q=e^{-i\frac{2\pi}{p'}},\quad
p'=2,3,\ldots.
\label{ppp}
\end{equation}
At $p'=2$, relations (\ref{algp}), (\ref{ppp}) 
reproduce equations (\ref{antir}).  
Using these relations, one finds that
operators $a^{p'}$ and  $\bar{a}{}^{p'}$ commute with operators $a$,
$\bar{a}$ and $P$. Hence, in irreducible representation they are
reduced to some constants. Assuming the existence of the vacuum
state $|0\rangle$, $a|0\rangle=0$, one finds that in Fock representation of
algebra (\ref{algp}), (\ref{ppp}), we have the relations
$
a^{p'}=\bar{a}{}^{p'}=0.
$
Multiplying the relation (\ref{algp}) from the left 
by the operator $P^{-1}=P^{p'-1}$,
we reduce it to the form of Lie-admissible algebra \cite{guon}:
$aT\bar{a}-\bar{a}Sa=1,$
$T=q^{-1}P^{-1},$
$
S=qP^{-1}.
$
Defining new creation-annihilation operators as
$
c=q^{-1/2}aP^{-1/2},
$
$
\bar{c}=q^{-1/2}P^{-1/2}\bar{a},
$
one gets finally the normalized 
$q$-deformed Heisenberg algebra \cite{qdef,macbi},
$
c\bar{c}-q\bar{c}c=1.
$
Let us note that the $q$-deformed Heisenberg algebra
in this form with $q$ being a primitive root of unity,
$q^{p'}=1$, was used by Chou for introducing
the so called genon statistics \cite{genon}.

Thus, the normalized form of
$R$-deformed Heisenberg algebra has the structure
of guon algebra \cite{guon}.
In the limit case $|\nu|\rightarrow\infty$
such guon-like algebra turns into the standard fermionic algebra
represented in the form of $R$-algebra (\ref{antir}). 
The natural generalization
of the latter into the algebra
(\ref{algp}), (\ref{ppp}) involving the phase operator $P$
results in $q$-deformed Heisenberg algebra with $q$ being
a primitive root of unity.

\section{Outlook}

Let us discuss briefly possible physical applications 
of the obtained results.

Recently it was shown \cite{deb} that finite-dimensional
representations of the $q$-deformed Heisenberg algebra give a
possibility to construct new variants of parasupersymmetry.
Their physical properties can be essentially different from the
properties of the parasupersymmetries realized in terms of the
standard parafermionic algebra \cite{rubspi,bede}.  It seems to
be interesting to investigate analogous possibility of
constructing parasupersymmetric systems with the help
of deformed parafermionic algebra obtained here.  Perhaps, the
intrinsic $Z_2$-grading structure peculiar to the deformation
could find physically interesting consequences.

We have established a universality of quadratic realization of 
$so(2,1)$ generators in terms of $R$-deformed Heisenberg algebra
operators $a^+$, $a^-$. 
The generators have the same operator form in the cases of 
infinite-dimensional half-bounded unitary representations and 
finite-dimensional non-unitary representations of (2+1)-dimensional
Lorentz group. 
This universality can be used for the construction of the 
set of linear differential equations which will describe either 
fractional spin fields (anyons) or ordinary integer and half-integer 
spin fields. 
The equations and corresponding field action will be governed
by the choice of the appropriate value of the deformation
parameter $\nu$ in 
the underlying $R$-deformed Heisenberg algebra \cite{plpre}.
Moreover, the related $osp(1|2)$ superalgebraic structure 
can supply us with a natural basis for realizing 
(2+1)-dimensional supersymmetry \cite{plpre}. 
This possibility is based on the fact that every described
$osp(1|2)$ representation contains a direct sum of two corresponding
$so(2,1)$ representations with spin shifted in one-half.

The constructed guon-like algebra of the form (\ref{1nu1}) or
(\ref{rfer})  contains the operator-valued
function $g$. Unlike the original guon algebra \cite{guon},
here $g^2\neq 1$ and $[g,c^\pm]\neq 0$.  The condition of the
form $[g,c^\pm]=0$ appeared in \cite{guon} from the requirement
of micro causality under assumption that observables should be
bilinear in fields or in creation-annihilation
operators.  On the other hand, it is known that in the
field-theoretical anyonic constructions involving the
Chern-Simons gauge field, there are observables (e.g., total
angular momentum operator) which are not bilinear in
creation-annihiliation operators \cite{ger}.  Moreover, the
gauge-invariant fields carrying fractional spin and statistics
themselves turn out to be nonlocal operators
\cite{ban} being decomposable in some infinite series 
in degrees of creation-annihilation operators
of the initial matter field. 
It seems that guon-like algebra appeared here
could also find some applications in the theory of anyons.
In particular, it could be useful for
establishing spin-statistics relation for fractional spin
fields within the framework of the group-theoretical
approach to anyons \cite{bosany}.

\vskip0.5cm
{\bf Acknowledgements}

I thank the Department of Physics of the Federal
University of Juiz de Fora,  where the part of this work
was done, for hospitality.
I am grateful to M.V. Cougo-Pinto for useful discussions.
%\newpage
\appendix

\section{Realization of R-para\-gras\-smann al\-geb\-ra 
on $q$-para\-gras\-smann algebra}
Due to the isomorphism between non-degenerate paragrassmann algebras
of the same order \cite{fik2},
$R$-paragrassmann algebra (\ref{aaa}),
(\ref{parag}) can be realized
with the help of $q$-paragrassmann algebra of order $2p+1$
\cite{fik1,fik2}.
The latter is given by the relations
\[
\partial \theta-q \theta\partial =1,\quad
\theta^{2p+1}=\partial^{2p+1}=0,\quad 
q^{2p+1}=1,\quad
\partial 1=0.
\]
For this algebra we have the relations
\[
\partial^n\theta=(n-1)_q\partial^{n-1}+q^n\theta\partial^n,\quad
\partial\theta^n=(n-1)_q\theta^{n-1}+q^n\theta^n\partial,
\]
where $n=1,\ldots,2p$,
and we have used the notation $(n)_q=\sum_{k=0}^n q^k$.
With the help of this algebra,
the operators $a^+$ and $R$
can be realized as $a^+=\theta$ and
\begin{equation}
R=\sum_{n=0}^{2p}f_n\theta^n\partial^n,
\label{r-q}
\end{equation}
where 
\[
f_0=1,\quad f_{n+1}=(-1)^{n+1}\frac{\prod_{l=0}^{n}(1+q^l)}
{\prod_{k=0}^n(k)_q},\quad
n=0,\ldots,2p-1,
\]
and so, $f_1=-2$ and $f_2=2$
for any $p$.
The operator $R$ anticommutes with $\theta$ and 
$q$-derivative $\partial$.
Our differentiation operator $a^-$
is realized in the form
\begin{equation}
a^-=\sum_{n=0}^{2p-1}g_n\theta^n\partial^{n+1}
\label{agq}
\end{equation}
with coefficients given by the recursive relations
\[
g_n(n)_q+g_{n-1}(q^n-1)=-(2p+1)f_n,\quad 
n=1,\ldots,2p-1,\quad g_0=-2p.
\]
Here the highest coefficient $g_{2p-1}$ has the form
$g_{2p-1}=(2p-1)f_{2p}/(1-q^{2p})$.

In the simplest case of $p=1$, one has
\[
a^-=-2\partial+6(1-q^2)^{-1}\theta\partial^2,\quad
R=1-2\theta\partial+2\theta^2\partial^2.
\]
{}For $p=2$, the operators $R$ and $a^-$ are given by
relations (\ref{r-q}) and (\ref{agq}) with nontrivial
coefficients 
\[
f_0=1,\quad -f_1=f_2=2,\quad
f_3=2q^2(1+q^2)\cdot(1+q)^{-1},\quad
f_4=2(3+2q^2+2q^3),
\]
\[
g_0=-2,\quad 
g_1=(6+4q)(1)_q^{-1},\quad
g_2=-4+2q(2)_q^{-1},\quad 
g_3=10(2q^3+q^2+q+2)(1-q^2)^{-1}.
\]

\section{Finite-dimensional matrix representations: $p=1,2$} 
Here we give the explicit expressions for matrix
finite-dimensional representations of the $R$-deformed
Heisenberg algebra and realization of $so(2,1)$ generators for
two simplest cases, $p=1$ and $p=2$.

For $p=1$, the corresponding operators are realized as
$$
a^+=\pmatrix{.  & .  & .\cr
    \sqrt{2} & . & .\cr
    . & \sqrt{2} & . \cr},\quad
a^-=\pmatrix{.  & -\sqrt{2}  & .\cr
             .  & . & \sqrt{2}\cr
    . & . & . \cr},
$$
where dots correspond to zero elements,
and $R=diag(1,-1,1)$, $\eta=diag(1,-1,-1)$,
${\cal J}_0=diag(-\frac{1}{2},0,\frac{1}{2})$,
$$
{\cal J}_1=\frac{1}{2}\pmatrix{.  & .  & -1\cr
    . & . & .\cr
    1 & . & . \cr},\quad
{\cal J}_2=\frac{i}{2}\pmatrix{.  & .  & -1\cr
             .  & . & . \cr
    -1 & . & . \cr}.
$$
Here generators ${\cal J}_{1,2}$ are related to 
${\cal J}_\pm$ as ${\cal J}_\pm={\cal J}_1\pm i{\cal J}_2$.
In the even subspace, where the operator $R$ takes the value
$+1$, the metric operator and generators of $so(2,1)$ are
reduced to the following matrices:
$\eta=\sigma_3$, ${\cal J}_0=-\frac{1}{2}\sigma_3$,
${\cal J}_1=-\frac{i}{2}\sigma_2$, 
${\cal J}_2=-\frac{i}{2}\sigma_1$.
Therefore, in this subspace the indefinite scalar product
is the Dirac scalar product: $(\Psi_1,\Psi_2)=\bar{\Psi}_{1}  
\Psi_{2}$, where $\bar{\Psi}=\Psi^\dagger\gamma_0$
is the Dirac conjugate 
wave function and $\gamma_0=-2{\cal J}_0=\sigma_3$.

For $p=2$ we have
$$
a^+=\pmatrix{.  & .  & . & . & . \cr
             2  & .  & . & . & .\cr
              . & \sqrt{2} & . & . & .\cr
             . & . & \sqrt{2} & . & .\cr
             . & . & . &  2& .\cr},\quad
a^-=\pmatrix{.  & -2& .& . & .\cr
             . & . & \sqrt{2} & . &.\cr
             . & . & . & -\sqrt{2} & .\cr  
             .  & . & . & . &2\cr
             .  & . & . & . & .\cr},
$$
and 
$R=diag(1,-1,1,-1,1)$, 
$\eta=diag(1,-1,-1,1,1)$,
${\cal J}_0=diag(-1,-\frac{1}{2},0,\frac{1}{2},1)$,
$$
{\cal J}_1=\frac{1}{2}\pmatrix{.  & .  & -\sqrt{2} & . & . \cr
             .  & .  & . & -1 & .\cr
              \sqrt{2} & . & . & . & -\sqrt{2}\cr
             . & 1 & . & . & .\cr
             . & . & \sqrt{2} & . & .\cr},\quad
{\cal J}_2=-\frac{i}{2}\pmatrix{.  & . & \sqrt{2} & . & .\cr
             . & . & . & 1 & .\cr
             \sqrt{2} & . & . & . & \sqrt{2} \cr  
             .  & 1 & . & . & .\cr
             .  & . & \sqrt{2} & . & .\cr}.
$$
On the even subspace the metric operator is reduced to
$\eta=diag(1,-1,1)$. This operator and corresponding $so(2,1)$
generators being reduced to this subspace
are unitary equivalent to the standard vector realization
of $(2+1)$-dimensional Lorentz group.
The latter is 
given by $\tilde{\eta}_{\alpha\beta}=diag(-1,1,1)$
and
$(\tilde{{\cal J}}_\mu)^{\alpha}{}_{\beta}
=-i\epsilon^{\alpha}{}_{\mu\beta}$,
$\alpha,\beta,\mu=0,1,2,$
with totally antisymmetric tensor normalized as
$\epsilon^{012}=1$ \cite{corply}.
The equivalence is established by the
relations $\tilde{\eta}=U\eta U^\dagger$,
$\tilde{{\cal J}}_\mu=U{\cal J}_\mu U^\dagger$,
where $U$ is the unitary matrix, 
$$
U=\frac{1}{\sqrt{2}}\pmatrix{.  & -\sqrt{2} & . \cr
              1 & . & 1 \cr
              -i& . & i\cr}.
$$

\end{document}